\documentstyle[prb,aps,graphicx,floats]{revtex}
\begin{document}
%\bibliographystyle{unsrt}
%\wideabs{
\draft
\title{ Theory of resonant Raman scattering of tetrahedral amorphous carbon}
\author{Mickael Profeta and Francesco Mauri} 
\address{Laboratoire de Min\'{e}ralogie-Cristallographie de Paris,\\ 
  Universit\'{e} Pierre et Marie Curie, 4 Place Jussieu, 75252, 
  Paris, Cedex 05, France}
\date{\today}
\maketitle
\begin{abstract}
We present a practical method to compute the
vibrational resonant Raman spectra in solids with delocalized excitations. 
We apply this approach to the study of tetrahedrous amorphous carbon.
We determine the vibrational eigenmodes and eigenvalues
using density functional theory in the local density approximation,
and the Raman intensities using a tight binding approximation.
The computed spectra are in good agreement with the experimental ones
measured with visible and UV lasers. 
We analyze the Raman spectra in terms of vibrational modes of
microscopic units.
We show that, at any frequency, the spectra are dominated by
the stretching vibrations.
We identify a very rapid inversion in the relative Raman intensities of
the $sp^2$ and $sp^3$ carbon sites with the frequency of the incident laser 
beam.
In particular, the spectra are dominated by $sp^2$ atoms below 4 eV and
by $sp^3$ atoms above 6 eV.
\end{abstract}

%78.30    Infrared and Raman spectra
%71.15.-m Methods of electronic structure calculations
%71.15.Mb Density functional theory, local density approximation
%63.50.+x Vibrational states in disordered systems
%}

\section{Introduction}

Tetrahedral amorphous carbon is a very studied material because of its physical
properties, which are close to those of diamond. 
It is composed of carbon atoms with $sp^3$ and $sp^2$ hybridization,
with a fraction of $sp^3$ sites larger than 40 \%.
The properties of tetrahedral amorphous carbon are very dependent 
on its microscopic structure. Thus, the
knowledge of the microscopic structure can be used to improve the synthesis of
the material.
Raman spectroscopy have often been applied to the study of 
tetrahedral amorphous carbon.\cite{Robertson} However visible Raman
spectroscopy is much more sensitive to $sp^2$ than to $sp^3$ 
carbons.\cite{visible}
Recently, it has been shown that it is possible to probe the vibrations 
of $sp^3$ carbons with UV Raman,\cite{uv,uv2} but few UV spectra have been
presented in literature  up to now.
In absence of further experimental data, a theoretical study can elucidate 
the detailed dependence of the Raman spectra on the laser frequency.

A theory to compute the Raman spectra under {\em non-resonant} conditions 
is well established.
Indeed, the Placzek approximation\cite{bruesh} links the
scattered Raman intensity with the electronic susceptibility of the
sample. However, for resonant Raman the Placzek approximation is not 
justified.
In this work, we develop a practical method for the calculation of 
{\em resonant} Raman spectra in solids with delocalized excitations.
We apply this theory to amorphous carbon, using density functional theory in the
local density approximation to compute the vibrational modes and frequencies 
and a tight binding approximation to obtain the resonant Raman intensities.
We compare our results to the experimental  
visible and UV Raman spectra and we analyze our theoretical spectra
in terms of the motion of microscopic units.

\section{Theoretical results}
\label{theory} 

In Raman spectroscopy a sample of matter is irradiated by
a monochromatic laser beam of pulsation $\omega_{\rm L}$, and the
intensity of the scattered light $I(\omega)$ is measured, as a function 
of the pulsation $\omega$.
The intensity $I(\omega)$ has a main elastic peak at 
the incident light pulsation, $\omega=\omega_{\rm L}$, and a smaller 
inelastic contribution at $\omega\ne\omega_{\rm L}$ associated to 
energy transfers between the light and the sample.
In this paper, we will focus on first-order vibrational Stokes spectroscopy, 
in which the light excites a single vibrational mode.

In order to describe the vibrational Raman scattering and to fix the
notation, we need to introduce the Born-Oppenheimer (BO) approximation.
Within the BO approximation the eigenstates of the Hamiltonian with
eigenvalues $E_{k}^e$ can be written as $ |\psi_{e,{\bf
R}}\rangle \otimes |\nu_{e,k} \rangle $, where the kets $
|\psi_{e,{\bf R}}\rangle$ and $|\nu_{e,k} \rangle $ are defined
on the Hilbert space of the electronic and nuclear coordinates,
respectively.  The kets $|\psi_{e,{\bf R}}\rangle$ are the
eigenvectors with eigenvalues $E_e({\bf R})$ of the electronic
Hamiltonian $H^{el}({\bf R})$, which depends parametrically on the
positions ${\bf R}$, of the N nuclei (${\bf R}$ represents a vector of
3N coordinates). For every $e$, each $E_e({\bf R})$ defines a BO
potential energy surface.  Finally the wave-functions
$|\nu_{e,k} \rangle $ satisfy the eigenvalue  equation:
\begin{equation} 
\label{bo}
[T + E_e({\bf R})] |\nu_{e,k} \rangle = E_{k}^e
|\nu_{e,k} \rangle ,
\end{equation}
where  $T$ is the kinetic energy of the  nuclei.
It is possible to define a vibrational energy as:
\begin{equation} 
\epsilon_{k}^e =  E_{k}^e - E_e({\bf R}^{\rm eq}_e),
\end{equation}
where ${\bf R}^{\rm eq}_e$ is the equilibrium position of the nuclei
on the BO surface $E_e({\bf R})$. 

In vibrational Raman spectroscopy, before the scattering, the system
is on the ground state BO surface, $E_0({\bf R})$, and in the vibrational
state $\epsilon_{i}^0$, after the scattering the system is still on
the ground state BO surface but in a different vibrational
state $\epsilon_{f}^0$. In first-order Stokes Raman the harmonic 
approximation is assumed and the final state differs from the initial one by 
the creation of a phonon with energy $\omega_s$.\cite{footnotehbar}

In non-resonant Raman scattering the energy of incident light
is very far from any electronic excitation.
Under this condition, a simple and well established theory to compute the 
scattering intensity $I(\omega)$, have been developed by 
Placzek.\cite{Placzek,bruesh}
Within the Placzek's approximation the intensity of scattered light
is expressed in terms of the electronic susceptibility at the 
laser frequency $\chi_{{\bf u}_{\rm L}{\bf u}}(\omega_{\rm L})$:
\begin{equation}
\label{placzek}
I({\bf u}_{\rm L},{\bf u},\omega) \propto \sum_s \frac{\omega^4}{\omega_s}
\left|\frac{\partial \chi_{{\bf u}_{\rm L}{\bf
u}}(\omega_{\rm L})}{\partial Q_s}\right|^2 (\langle n_{s} \rangle +
1) \delta (\omega -\omega_{\rm L} + 
\omega_s),
\end{equation}
where ${\bf u}_{\rm L}$ and  ${\bf u}$ are the polarization of the 
incident and scattered light, $\langle n_{s} \rangle =
[\exp(\omega_s/(k_BT))-1]^{-1}$ is the average thermal occupation of the 
vibrational state $s$, and
the derivative of the
susceptibility with respect to the vibrational normal mode $Q_s$ is:
\begin{equation}
\frac{\partial \chi_{{\bf u}_{\rm L}{\bf u}}(\omega_{\rm L})}{\partial
  Q_s}=\sum_J 
\sum_{\alpha=x,y,z} \frac{\partial \chi_{{\bf u}_{\rm L}{\bf
  u}}(\omega_{\rm L})}{\partial 
R_J^\alpha}q_J^{\alpha,s} = \frac{\partial \chi_{{\bf u}_{\rm L}{\bf
u}}(\omega_{\rm L})}{\partial {\bf R}}\cdot{\bf q}^s.
\end{equation}
Here $ q_J^{\alpha,s}$ is a component of the unit vector of normal
mode $Q_s$ for vibration $s$ on the atom J with normalization
${\bf q}^s \cdot {\bf q}^{s^\prime}=\delta_{s,s^\prime} $.
The  Placzek's approximation have been used successfully for the
first principles computation of the non-resonant vibrational Raman intensity,
(see e.g. Refs. \onlinecite{porezag,giannozzi}).

In resonant Raman spectroscopy the energy of incident light
is in resonance with an electronic transition.
Under this condition the Placzek's approximation is, in principle,
not justified. However Eq.(\ref{placzek}) has also been applied to
resonant Raman scattering in solids\cite{cardona} in order to interpret 
experimental measurements on semiconductors.

In this paper, we will derive an expression for the scattering intensity in 
solids
which is valid both in the non-resonant and resonant case.
This derivation will be used in the following section 
to analyze the Raman spectra of amorphous carbon.

We start our derivation from the general expression for the vibrational Raman 
intensity, which can be obtained from theory of second quantization of 
light:\cite{quantique}
\begin{eqnarray}
I({\bf u}_{\rm L},{\bf u},\omega) & \propto & \sum_{i,f} \rho(i)
\omega^4 |{\mathcal{A}}_{i,f}|^2\delta(\omega-\omega_{\rm
  L}-\epsilon_i^0+\epsilon_{f}^0), 
\end{eqnarray}
with
\begin{eqnarray}
\label{initial}
\lefteqn{{\mathcal{A}}_{i,f}=\sum_{e,k} \left( \frac{ \langle
      \nu_{0,i}|f^e_{{\bf u}_{\rm L}}({\bf R})|\nu_{e,k}\rangle\
      \langle \nu_{e,k}|f_{{\bf u}}^{e\ast}({\bf R})|\nu_{0,f}\rangle}
    {E_e({\bf R}^{\rm eq}_e)+\epsilon_{k}^e -E_0({\bf R}^{\rm
        eq}_0)+\epsilon_{i}^0-\omega_{\rm L}-i\gamma}\right.}\nonumber \\ 
  && \left.+ \frac { \langle \nu_{0,i}| f^e_{{\bf u}}({\bf
        R})|\nu_{e,k}\rangle \ \langle\nu_{e,k}|f_{{\bf u}_{\rm
          L}}^{e\ast}({\bf R})|\nu_{0,f} \rangle } 
    {E_e({\bf R}^{\rm eq}_e)+\epsilon_{k}^e -E_0({\bf R}^{\rm
        eq}_0)+\epsilon_{i}^0+\omega+i\gamma}\right). 
\end{eqnarray}
Here, $\rho(i)$ is the probability to find the system in the initial state $i$,
$\gamma$ is a small real number added for calculation
in order to treat correctly the poles of the expression and
\begin{equation}
f^e_{\bf u}({\bf R})=\frac{1}{\sqrt{V}}\langle \psi_{0,{\bf R}}|{\bf u}\cdot{\bf D}({\bf
r},{\bf R})|\psi_{e,{\bf R}} \rangle,
\end{equation}
where ${\bf D}$ is the ionic and electronic dipole moment, $V$ is the sample
volume, and ${\bf r}$ is the position vector for the electrons.
Placzek's expression, Eq.~(\ref{placzek}), has been derived 
in the non resonant case starting 
from Eq.~(\ref{initial}).\cite{Placzek,bruesh}
Notice that Eq.~(\ref{initial}) is much more complicated to be evaluated than
the Placzek's expression, Eq.~(\ref{placzek}),
since it requires the knowledge
of the vibrational eigenfunctions, $|\nu_{e,k}\rangle$, on the
excited BO surfaces.

Before presenting our derivation we recall, for comparison, that of the 
Placzek's approximation.\cite{Placzek,bruesh} In the Placzek derivation Eq.~(\ref{initial})
is approximated by: 
\begin{eqnarray}
\label{plac_1}
  \lefteqn{{\mathcal{A}}_{i,f}=\sum_{e} \left( \frac{\sum_{k} \langle
        \nu_{0,i}|f^e_{{\bf u}_{\rm L}}({\bf R})|\nu_{e,k}\rangle\
        \langle \nu_{e,k}|f_{{\bf u}}^{e\ast}({\bf R})|\nu_{0,f}\rangle}
      {E_e({\bf R}) -E_0({\bf R})-\omega_{\rm
          L}-i\gamma}\right.}\nonumber \\  
  && \left.+ \frac { \sum_{k} \langle \nu_{0,i}| f^e_{{\bf
          u}}({\bf R})|\nu_{e,k}\rangle \ \langle\nu_{e,k}|f_{{\bf u}_{\rm 
          L}}^{e\ast}({\bf R})|\nu_{0,f} \rangle } 
    {E_e({\bf R}) -E_0({\bf R})+\omega+i\gamma}\right).
\end{eqnarray}
Using the completeness relation $\sum_k
|\nu_{e,k}\rangle\langle \nu_{e,k}|=1$, Eq.~(\ref{plac_1}) becomes:
\begin{equation}
\label{plac_2}
{\mathcal{A}}_{i,f}=\sum_{e} \left(\frac{ \langle
    \nu_{0,i}|f^e_{{\bf u}_{\rm L}}({\bf R})f_{{\bf
        u}}^{e\ast}({\bf R})|\nu_{0,f}\rangle} 
  {E_e({\bf R}) -E_0({\bf R})-\omega_{\rm
    L}-i\gamma} 
   + \frac { \langle \nu_{0,i}| f^e_{{\bf
          u}}({\bf R})f_{{\bf u}_{\rm 
          L}}^{e\ast}({\bf R})|\nu_{0,f} \rangle } 
    {E_e({\bf R}) -E_0({\bf R})+\omega+i\gamma}\right),
\end{equation}
which finally leads to Eq.~(\ref{placzek}).
In order to obtain Eq.~(\ref{plac_1}), we drop the dependence of
the denominators of Eq.~(\ref{initial}) on the excited 
vibrational states $|\nu_{e,k}\rangle$, by replacing $[E_e({\bf R}^{\rm
  eq}_e)+\epsilon_{k}^e -E_0({\bf R}^{\rm eq}_0)+\epsilon_{i}^0]$ with
$[E_e({\bf R}) -E_0({\bf R})]$.
In this way, we neglect in the denominators an energy term 
\begin{equation}
\Delta_e^k=E_e({\bf R}^{\rm eq}_e)- E_e({\bf
  R})+\epsilon_{k}^e-[E_0({\bf R}^{\rm eq}_0)-E_0({\bf
  R})+\epsilon_{i}^0].
\end{equation}
This approximation is justified if $E_e({\bf
  R}) -E_0({\bf R})-\omega_{\rm L} \gg \Delta_e^k$. Since we are
using the completeness relation, the
above inequality should, in principle, held for every vibrational
states $k$, and not 
just for the one phonon excitations.  
In particular, the relation must be verified for excitations, and
  energy differences $\Delta_e^k$, involving
an arbitrary number of phonons and even for unbounded states in
the energy continuum. This condition can obviously never be
satisfied. The use of Placzek approximation relies on the hope that
the inclusion of excitations involving a limited amount of phonons are
sufficient to fulfill the completeness relation.
When the laser frequency is close to an electronic transition, the
Placzek's approximation could  not be applied in general since, in
this case, the 
dependence on $\Delta_e^k$ could not be ignored and the
completeness relation could not be used.

Here, to compute the  Raman spectra in a 
periodic solid, we simplify Eq.~(\ref{initial}) using an approximation 
different from Placzek's.
In a periodic solid with dispersive bands, the electronic excitations are 
generally delocalized. In this case, the excited BO surfaces are
obtained from the ground-state one by a constant vertical
displacement, and we can assume that 
$|\nu_{e,k}\rangle  \simeq |\nu_{0,k}\rangle $, 
${\bf R}^{\rm eq}_e \simeq {\bf R}^{\rm eq}_0$,
and $\epsilon_{k}^e \simeq
\epsilon_{k}^0 $. This leads to:
\begin{eqnarray}
\label{A}
\lefteqn{ {\mathcal{A}}_{i,f}({\bf u}_{\rm L},{\bf u})=\sum_{e,k}
\left( \frac{ \langle \nu_{0,i}|f^e_{{\bf u}_{\rm L}}({\bf R})|\nu_{0,k}\rangle
\ \langle \nu_{0,k}| f_{\bf u}^{e\ast}({\bf R})|\nu_{0,f}\rangle
}{ E_e({\bf R}^{\rm eq}_0)-E_0({\bf R}^{\rm
eq}_0)+\epsilon_{k}^0-\epsilon_{i}^0-\omega_{\rm L}-i\gamma}\right.}
\nonumber \\ && \left. + \frac{ \langle \nu_{0,i}|f^e_{\bf u}({\bf
R})|\nu_{0,k}\rangle \ \langle \nu_{0,k}|f_{{\bf u}_{\rm L}}^{e\ast}({\bf
R})|\nu_{0,f}\rangle }{ E_e({\bf R}^{\rm eq}_0)-E_0({\bf R}^{\rm
eq}_0)+\epsilon_{k}^0-\epsilon_{i}^0+\omega+i\gamma}\right).
\end{eqnarray}
We expand the expression of $f^e_{{\bf u}}({\bf R})$ at first order
around ${\bf R}^{\rm eq}_0$,  $f^e_{{\bf u}}({\bf R}) \simeq f^e_{{\bf u}}({\bf
R}^{\rm eq}_0) +({\bf R}-{\bf R}^{\rm eq}_0)\cdot  \partial f^e_{\bf
u}({\bf R}^{\rm eq}_0)/\partial {\bf R}$ and we extract all the terms
of degree one in $({\bf R}-{\bf R}^{\rm eq}_0)$ in 
Eq.~(\ref{A}). The term ${\bf R}-{\bf R}^{\rm eq}_0$ can be written
using the operators of creation and annihilation of one phonon of
energy $\omega_s$. If we extract Stokes terms, we can write the amplitude as 
a function of a vibration $s$, 
\begin{eqnarray}
\label{sumovers}
I({\bf u}_{\rm L},{\bf u},\omega) & \propto & \sum_{s,n_s}\rho(n_s)\omega^4 
|{\mathcal{A}}_{s}|^2\delta(\omega-\omega_{\rm L}+\omega_s),
\end{eqnarray}
where $n_s$ is the occupation number of the mode $s$, 
$\omega_s=\epsilon_f^0-\epsilon_{i}^0$, and
\begin{eqnarray}
\label{as}
{\mathcal{A}}_{s}&=&\sum_e 
\sqrt{\frac{n_s+1}{2\omega_s}}
\left( \frac{
{\bf q}_s \cdot  
(\frac{\partial f^e_{{\bf u}_{\rm L}}}
{\partial {\bf R}})_{{\bf R}^{\rm eq}_0}  
f_{\bf u}^{e\ast}({\bf R}^{\rm eq}_0) }
{ E_e({\bf R}^{\rm eq}_0)-E_0({\bf R}^{\rm eq}_0)
+\omega_s-\omega_{\rm L}-i\gamma}
+\frac{ 
f^e_{{\bf u}_{\rm L}}({\bf R}^{\rm eq}_0){\bf q}_s \cdot
(\frac{\partial f_{\bf u}^{e\ast}}{\partial {\bf R}})_{{\bf R}^{\rm eq}_0} }
{ E_e({\bf R}^{\rm eq}_0)-E_0({\bf R}^{\rm eq}_0)
-\omega_{\rm L}-i\gamma}\right. \\ 
&+&\frac{{\bf q}_s \cdot (\frac{\partial
f^e_{\bf u}}{\partial {\bf R}})_{{\bf R}^{\rm eq}_0} f_{{\bf
u}_{\rm L}}^{e\ast}({\bf R}^{\rm eq}_0) }{ E_e({\bf R}^{\rm eq}_0)-E_0({\bf
R}^{\rm eq}_0)+\omega_{\rm L}+i\gamma} \nonumber  \left. + \frac{ f^e_{\bf
u}({\bf R}^{\rm eq}_0){\bf q}_s \cdot(\frac{\partial f_{{\bf u}_{\rm L}}^{e\ast}}{\partial
{\bf R}})_{{\bf R}^{\rm eq}_0} }{ E_e({\bf R}^{\rm eq}_0)-E_0({\bf
R}^{\rm eq}_0)+\omega_{\rm L}+i\gamma-\omega_s}\right).
\end{eqnarray}
Finally, we neglect the dependence of the denominator over $\omega_s$. Indeed, 
we expect that ${\mathcal{A}}_s$ has the same behavior with respect to
$\omega_L$ as the electronic susceptibility, $\chi(\omega_L)$. In a solid
with delocalized excitations, away from the van Hove singularities,
$\chi(\omega_L)$ is a smooth function such that $\chi(\omega_L\pm
\omega_s)\simeq \chi(\omega_L)$. Therefore, we expect that in a solid with
delocalized excitations we can drop the term $\pm \omega_s$ in the
denominator of Eq.~(\ref{as}).
Notice that the validity requirements of our approximation are
less stringent than those of the Placzek's approximation.
Our approximation neglects the dependence of the denominators on
the one phonon excitation energies $\omega_s$, whereas the Placzek's 
approximation neglects the multi-phonons excitation  energies
$\Delta_e^k$. Therefore our approximation requires that
$\chi(\omega)$ is flat on the energy scale of one phonon excitations
whereas the Placzek's approximation requires that
$\chi(\omega)$ is flat on the energy scale of multiple phonon
excitations.

\begin{equation}
\label{kappa}
{\mathcal{A}}_{s}= \sqrt{
\frac{n_s+1}{2\omega_s}}\  {\bf q}_s\cdot
\left. \frac{\partial \kappa({\bf R},{\bf R}',\omega_{\rm L})}{\partial {\bf
R}}\right|_{{\bf R}={\bf R}'={\bf R}^{\rm eq}_0},
\end{equation}
where
\begin{equation}
\label{kappa1}
\kappa_{{\bf u}_{\rm L},{\bf u}}({\bf R},{\bf R}',\omega_{\rm L})=\sum_e
\frac{f^e_{{\bf u}_{\rm L}}({\bf R})f_{\bf u}^{e\ast}({\bf R})} {E_e({\bf
R}')-E_0({\bf R}')-\omega_{\rm L}-i\gamma}  + \frac{f^e_{\bf u}({\bf
R})f_{{\bf u}_{\rm L}}^{e\ast}({\bf R})} {E_e({\bf R}')-E_0({\bf
R}')+\omega_{\rm L}+i\gamma}.
\end{equation}
Eqs. (\ref{sumovers}), (\ref{kappa}), and (\ref{kappa1}) are the working 
expressions that we will use in the next section to compute the Raman 
intensity.

It is interesting to compare our final expression with that of the Placzek's 
approximation.
To this purpose we notice that the electronic susceptibility $\chi$ can be 
obtained as a special case of the function $\kappa$, 
indeed $\chi_{{\bf u}_{\rm L},{\bf u}}({\bf R},\omega_{\rm L})=
\kappa_{{\bf u}_{\rm L},{\bf u}}({\bf R},{\bf R},\omega_{\rm L})$.
Thus, the  Placzek's expression differs from the expression we derived just by
an additional partial derivative. Indeed, if we assume that 
\begin{equation}
{\mathcal{A}}_{s}^{\rm Placzek}= \sqrt{
\frac{n_s+1}{2\omega_s}}\  {\bf q}_s\cdot
\left[ \frac{\partial \kappa({\bf R},{\bf R}',\omega_{\rm L})}{\partial {\bf
R}}+\frac{\partial \kappa({\bf R},{\bf R}',\omega_{\rm L})}{\partial {\bf
R}'}\right]_{{\bf R}={\bf R}'={\bf R}^{\rm eq}_0},
\end{equation}
we obtain Eq.~(\ref{placzek}) from Eq.~(\ref{sumovers}).

\section{Calculation on amorphous carbon}

\begin{figure}[h]
\centerline{\includegraphics*[angle=0,width=8cm]{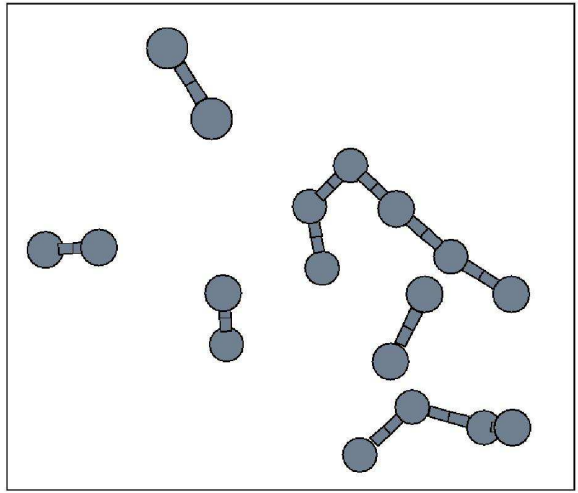}}
\caption{\label{sp2}
A ball and stick model of the $sp^2$ carbon atoms in the unit cell. 
A bond is drawn when the distance between two atoms
is less than 1.9 \AA. The $sp^2$   atoms are arranged in chains, no 
aromatic ring are present in the model.}
\end{figure}

We compute the vibrational spectra and the Raman intensities of
tetrahedral amorphous carbon using a model generated in Ref.
\onlinecite{drabold} by tight binding (TB) molecular dynamics.
The model contains 64 atoms in a periodic cubic cell of 7.57 \AA.
Among the 64 atoms, 18 atoms are three-fold coordinated ($sp^2$ hybridized) 
and the others are four-fold coordinated ($sp^3$ hybridized).
In this model all $sp^2$ atoms are inserted into a double bonded link,
i.e. there are no electronic defects associated to dangling $p$ orbitals. 
This kind of defects has been identified in other computed generated models of
amorphous carbon.\cite{drabold,mauriNMRac,bernasconi} Fig. \ref{sp2}
shows a ball and stick model of the  $sp^2$ carbon atoms in the
cell. The $sp^2$ atoms are arranged in four pairs and two chains of
four and six atoms respectively. No aromatic ring are present in the sample.  
We compute the dynamical matrix and the vibrational eigenvalues
and eigenvectors using density functional theory in the local density 
approximation.\cite{cpmd}
The C atoms are described by a Troullier Martins\cite{tm2} 
pseudo-potential with $s$ non-locality. 
The wave-functions are expanded in a plane wave basis set with a kinetic energy cut-off of 40 Ry. The dynamical
matrix is obtained by displacing each atom and computing the resulting
forces. The Brillouin zone is sampled with the $\Gamma$ ${\bf k}$-point only.
We compute the  intensities of the resonant Raman spectra using 
Eqs. (\ref{sumovers}), (\ref{kappa}), and (\ref{kappa1}).
The evaluation of the intensity at a resonant frequency requires a very 
fine Brillouin zone k-point mesh, for which a fully ab-initio calculation
is not affordable. 
Therefore we evaluate the function 
$\kappa_{{\bf u}_{\rm L},{\bf u}}({\bf R},{\bf R}',\omega_{\rm L})$ and
the electronic susceptibility 
$\chi_{{\bf u}_{\rm L},{\bf u}}({\bf R},\omega_{\rm L})$
within a TB approximation. We use a TB model with  {\it s} and
{\it p} orbitals.\cite{kmho} 
We verified that this TB Hamiltonian well reproduces the ab-initio
random-phase-approximation (RPA) dielectric 
constant of diamond\cite{rpa} for laser frequencies
smaller than 10 eV, which is the range of energies used in this work
for the calculation of the Raman spectra. The correct description of
the dielectric 
constant for frequencies larger than 10 eV would require the inclusion in
the TB model of {\it d} or {\it s$^*$} orbitals. 
The Brillouin zone of the 64 atoms supercell is sampled 
with 108 special k-points. 
We use a value of  0.05 eV for the constant $\gamma$ in Eq.~(\ref{kappa1}).
The derivative in Eq.~(\ref{kappa}) is computed by numerical differentiation.

\begin{figure}[h]
\centerline{\includegraphics*[angle=0,width=8cm]{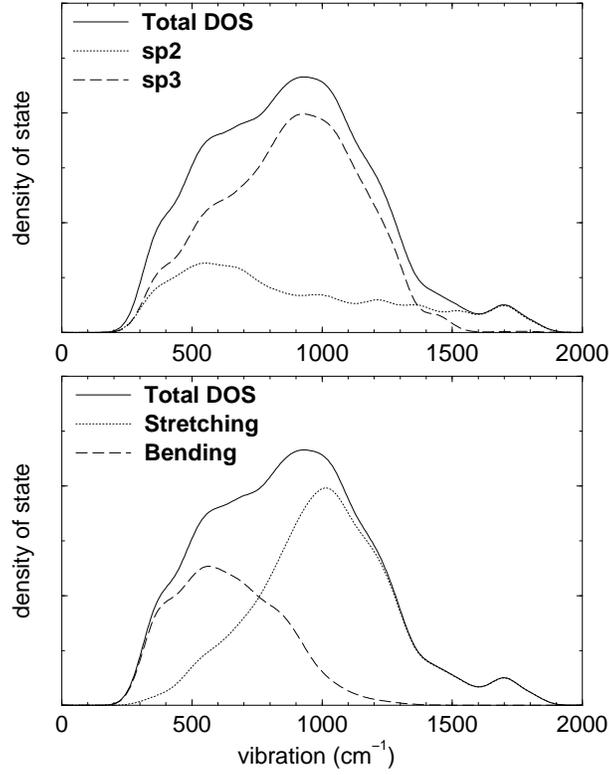}}
\caption{\label{dos} 
The total vibrational density of states of tetrahedral amorphous
carbon is decomposed 
in the contributions coming from $sp^2$ and $sp^3$ carbon sites (upper panel) 
and in the contributions coming from bending and stretching modes 
(lower panel). 
To obtain  continuous curves in this figure and in 
Figs. \ref{raman-vis} and \ref{raman-uv} we use a Gaussian smearing
with $\sigma=50$  $\mbox{cm}^{-1}$.} 
\end{figure}

In Fig.~\ref{dos} we present our theoretical vibrational density of 
state (DOS). 
The total DOS is decomposed in its partial contributions coming from the 
$sp^2$ and $sp^3$ hybridized carbons and from the stretching and bending 
modes.\cite{footnotestreben}
The $sp^2$ vibrations cover the entire range of frequencies of the
total DOS, whereas the $sp^3$ vibrations form a large peak centered at 
1000 $\mbox{cm}^{-1}$, which dies over 1400 $\mbox{cm}^{-1}$.
Thus, the modes above 1400 $\mbox{cm}^{-1}$ involve only $sp^2$ carbons.
The decomposition between bending and stretching  contributions
shows two separated peaks, centered at 600 $\mbox{cm}^{-1}$ and 
1000 $\mbox{cm}^{-1}$, respectively.
Above 1000 $\mbox{cm}^{-1}$ the stretching modes contribute alone to the DOS.
Similar theoretical results have been presented in  
Ref. \onlinecite{francesco} for a model of hydrogenated amorphous carbon.

\begin{figure}[h]
\centerline{\includegraphics*[angle=-90,width=8cm]{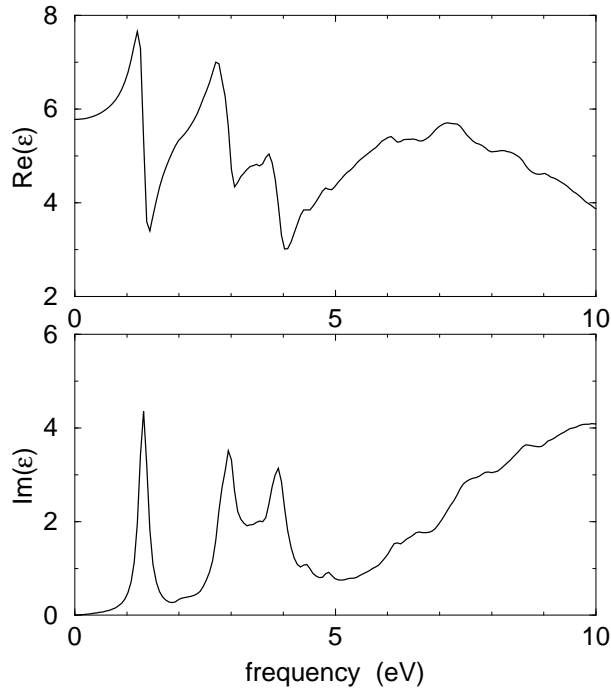}}
\caption{\label{susc} Real and imaginary parts of the electronic 
dielectric constant of tetrahedral amorphous carbon, 
$\epsilon(\omega_{\rm L})$, as a function of $\omega_{\rm L}$.}
\end{figure}

Before discussing the results for the Raman spectra, we present in 
Fig. \ref{susc} the real and imaginary part of the electronic 
dielectric constant, $\epsilon(\omega_L)=1+4\pi\chi(\omega_L)$. 
Our theoretical value of 5.8 for the static  dielectric constant
is in good agreement with the value of 6.2, measured experimentally
for a tetrahedral amorphous carbon sample.\cite{susceptibilite}
Notice that, for $\omega > 5$ eV, the function $\epsilon(\omega_L)$ is  
quite smooth, whereas, under 5 eV, three resonant peaks at 1.3, 
3.0 and 3.9 eV show up.
The presence of such peaks, which involve transitions between localized 
$\pi$ states, is an 
artifact due to the limited size of our model which contains only 
9 $\pi$-bonds. This kind of features should disappear in larger models. 
The electronic structure has a gap of about 0.9 eV, above this value the 
incident light can be absorbed by the sample, as shown by the non zero value 
of the imaginary part of the dielectric constant. Thus, Raman spectra 
both in the visible and ultra-violet (UV) regions are collected under resonant 
conditions. 
   
\begin{figure}[h] 
\centerline{\includegraphics*[angle=0,width=8cm]{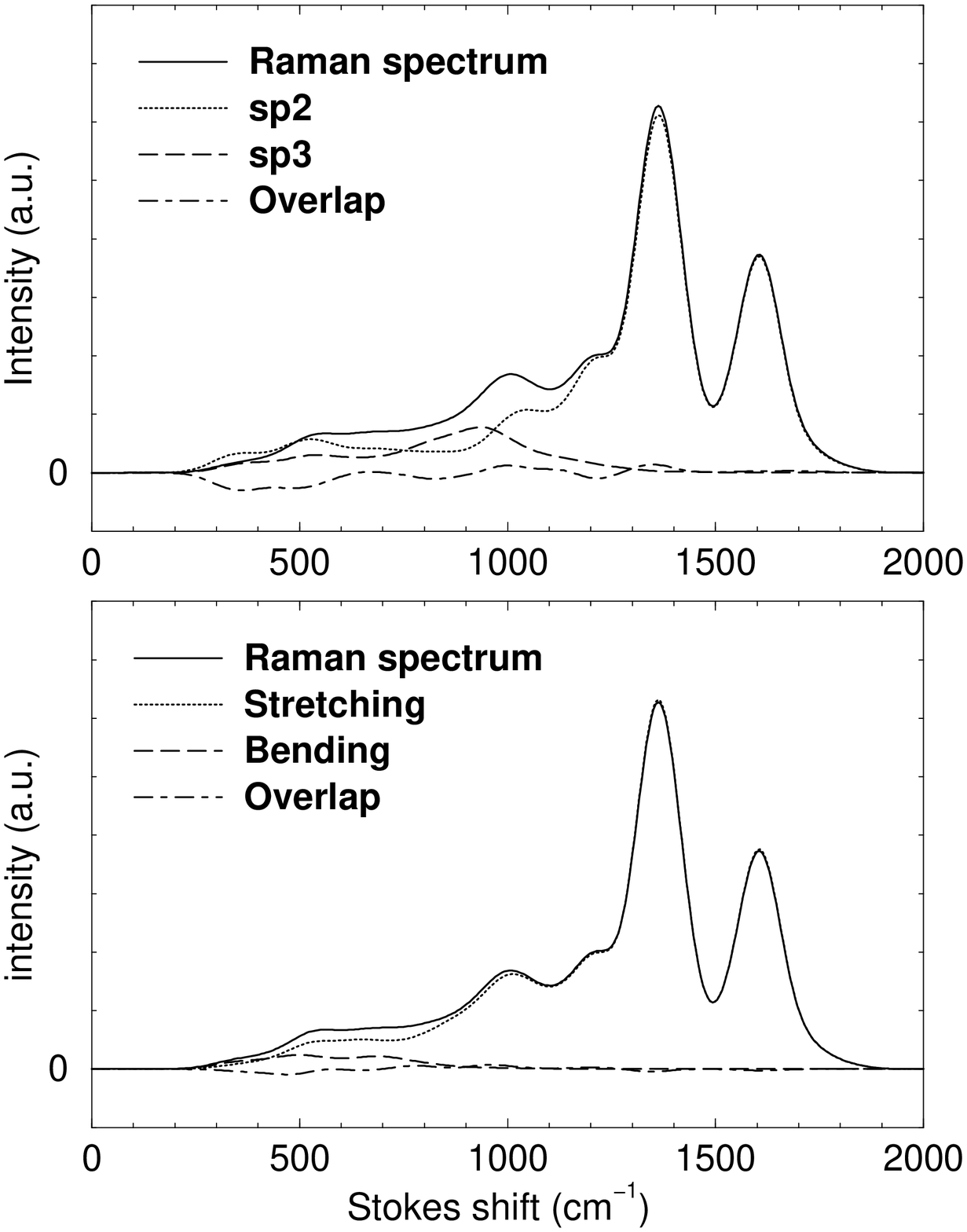}}
\caption{\label{raman-vis} 
Visible Raman spectrum of tetrahedral amorphous carbon 
computed for $\omega_{\rm L}= 1.8$ eV. The Raman intensity is plotted
as a function of the Stokes shift, $(\omega_{\rm L}-\omega)$.
The total Raman spectrum is decomposed in the contributions coming from $sp^2$,
 $sp^3$ atoms, and their overlap (upper panel)
and in the contributions coming from stretching and bending vibrations and 
their overlap (lower panel). Note that the presence of two peaks for
energies larger than 1200 cm$^{-1}$ is an artifact due to the small
total number of $sp^2$ sites in  our theoretical model.
}
\end{figure}

\begin{figure}[h]
\centerline{\includegraphics*[angle=0,width=8cm]{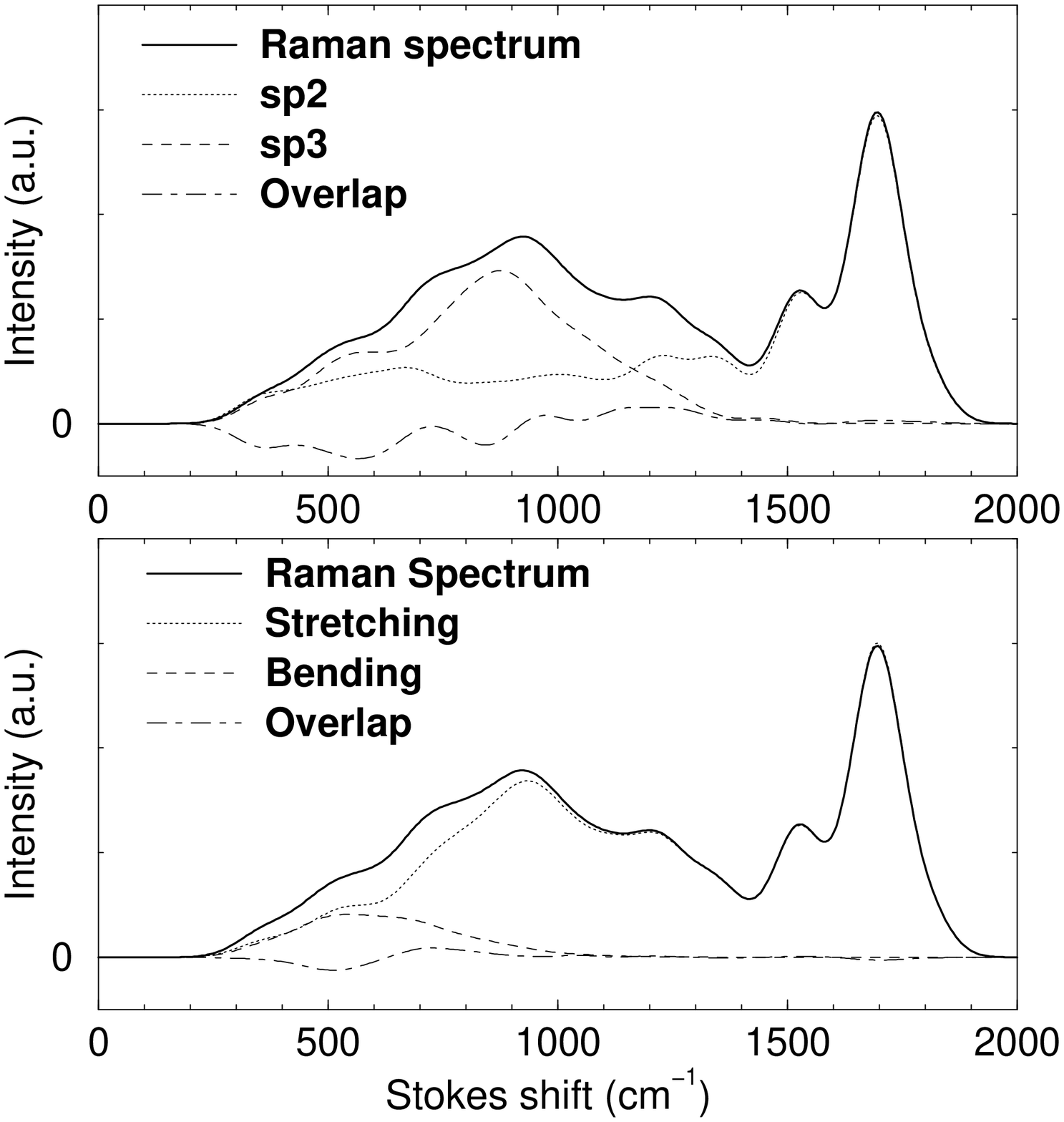}}
\caption{\label{raman-uv} 
UV Raman spectrum of tetrahedral amorphous carbon 
computed for $\omega_{\rm L}= 4.3$ eV. 
The Raman intensity is plotted
as a function of the Stokes shift, $(\omega_{\rm L}-\omega)$.
The total Raman spectrum is decomposed in the contributions coming from $sp^2$,
 $sp^3$ atoms, and their overlap (upper panel)
and in the contributions coming from stretching and bending vibrations and
their overlap (lower panel).
}
\end{figure}

Fig. \ref{raman-vis} and \ref{raman-uv} present the calculated Raman spectra of 
amorphous carbon for an incident laser beam of 1.8 eV and 4.3 eV, respectively.
The laser beam energies are chosen to be away from the three resonant peaks 
of $\epsilon(\omega_L)$.
For each spectrum, we show a decomposition between $sp^2$, $sp^3$,
bending, and stretching vibrations.
We also plot the overlap between the different contributions\cite{overlap}. The
decompositions are meaningful only if the overlap is small.
Again, the details of the shape of the Raman spectra coming from $sp^2$ atoms
are affected by the limited number of $sp^2$-$sp^2$ bonds in our model.
For example the division of the contribution above 1300 $\mbox{cm}^{-1}$ in 
two peaks is an artifact due to the lack of statistic.
In good agreement with the experimental observation,\cite{visible}
the visible Raman spectrum is essentially due to $sp^2$  carbons which
give a large contribution above 1200 $\mbox{cm}^{-1}$, centered at 1500 
$\mbox{cm}^{-1}$. Only a little $sp^3$ contribution, centered at 1000
$\mbox{cm}^{-1}$, can be noticed. 
In the UV Raman spectrum, on 
the contrary, a contribution of $sp^3$ hybridized atoms can clearly be noticed
through the large peak centered at 1000 $\mbox{cm}^{-1}$. Due to a larger 
number of $sp^3$ hybridized atoms in our model, 
the shape of this peak is more
reliable, and its position and shape closely match the experimental results.
\cite{uv,uv2} A peak centered at 1600  $\mbox{cm}^{-1}$, due to $sp^2$ sites, 
can still be observed, as in experiments.
Both visible and UV Raman spectra  are in a large 
majority due to the stretching vibrations, only a little contribution of 
bending vibrations can be noticed at about 600 $\mbox{cm}^{-1}$,
which has sometimes been observed in the experimental spectra (see e.g. Fig. 1
of Ref. \onlinecite{uv}).

\begin{figure}[h]
\centerline{\includegraphics*[angle=0,width=8cm]{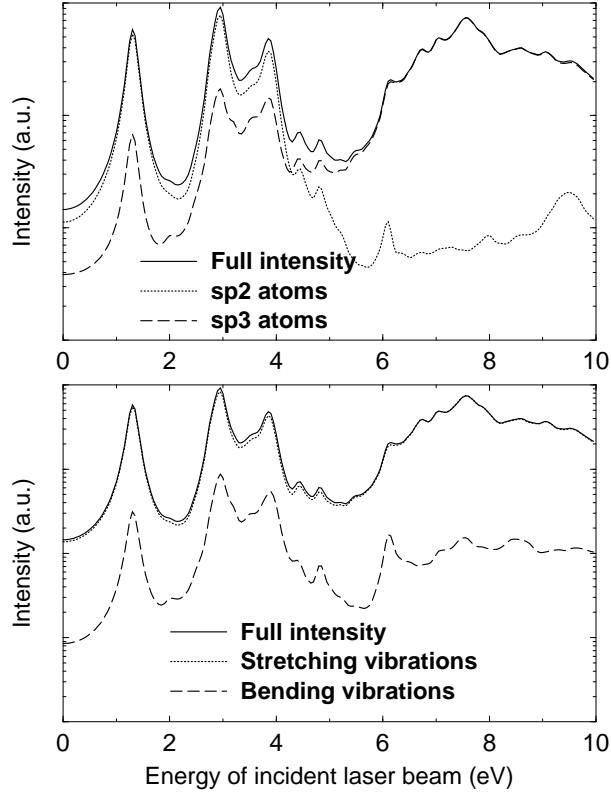}}
\caption{\label{intensity} 
Absolute integrated intensities of the Raman spectra of
tetrahedral amorphous carbon plotted as 
a function of the energy of incident laser beam, $\omega_{\rm L}$.  
In this figure, we present the absolute intensity divided by $\omega^4$ (see Eq.~(\ref{sumovers}))  of the 64 atoms model using 
a logarithmic scale for the intensity axis.
The total integrated intensity is decomposed in the contributions coming from
$sp^2$ and $sp^3$ atoms (upper panel)
and in the contributions coming from stretching and bending vibrations 
(lower panel).
}
\end{figure}

\begin{figure}[h]
\centerline{\includegraphics*[angle=0,width=8cm]{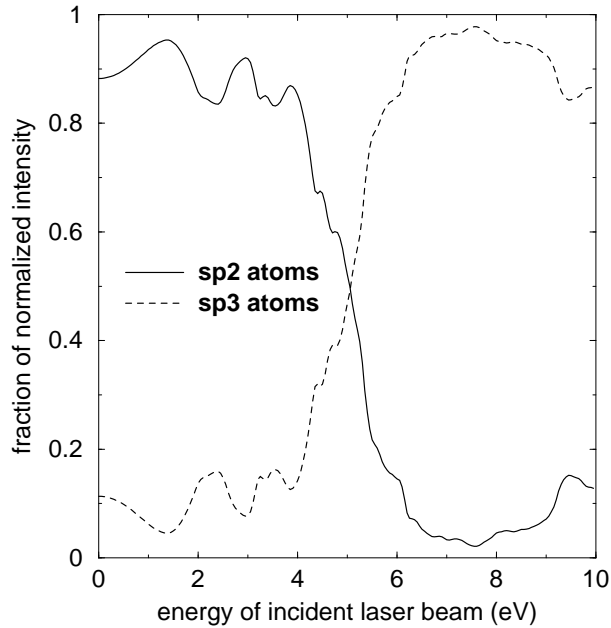}}
\caption{\label{int-rel}
Fractional integrated intensities of the Raman spectra,
plotted as a function of the energy of incident laser beam, $\omega_{\rm L}$.
Here, we present the fraction of the $sp^2$ and $sp^3$ intensity per 
site. 
For each value of $\omega_{\rm L}$, we compute the intensities per site,
$i^{sp3}(\omega_{\rm L})$ and $i^{sp2}(\omega_{\rm L})$, by dividing the 
absolute $sp^3$ and $sp^2$ intensities presented in Fig. \ref{intensity},
by the number of $sp^3$ and $sp^2$ atoms in the model, respectively.
In this figure we plot the fractional intensities,
$i^{sp3}(\omega_{\rm L})/[i^{sp3}(\omega_{\rm L})+i^{sp2}(\omega_{\rm L})]$ and
$i^{sp2}(\omega_{\rm L})/[i^{sp3}(\omega_{\rm L})+i^{sp2}(\omega_{\rm L})]$.
}
\end{figure}

For a given laser pulsation, $\omega_{\rm L}$, we compute the integrated Raman
Stokes intensities as the total area of the Stokes spectra.
The total integrated Raman intensity and its decomposition 
in $sp^2$, $sp^3$, bending, and stretching contributions are presented in 
Figs.  \ref{intensity} and \ref{int-rel} as a function of $\omega_{\rm L}$.
These data show that whatever the energy of the incident
light, the intensity of the Raman spectrum is always dominated by stretching
modes,  whose  contribution is always at least an order of magnitude larger
than that of bending modes.
Regarding the relative decomposition in the $sp^2$ and $sp^3$ contributions,
Fig.  \ref{intensity} and \ref{int-rel} clearly show an inversion at about
5 eV. Under 4 eV almost 90\% of the intensity is due  to $sp^2$
carbons, whereas over 6 eV almost 90\% of the intensity is due to $sp^3$
carbons.
Notice that this inversion happens very quickly,
thus, in an experimental measurement a little increase in the
energy of the incident light in the UV region can lead to a
major modification of the Raman spectra.

\begin{figure}[h]
\centerline{\includegraphics*[angle=-90,width=8cm]{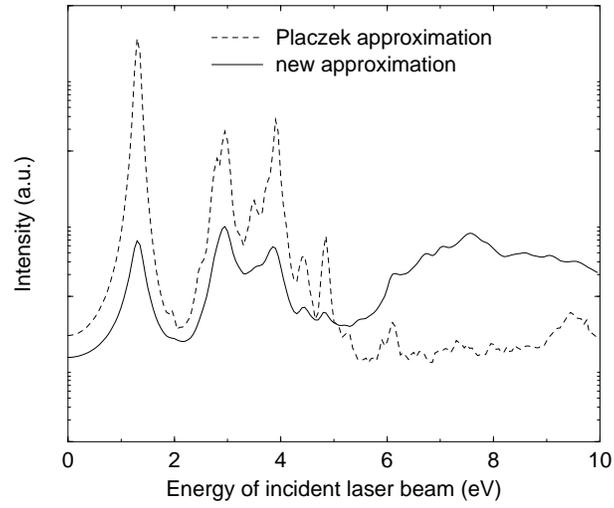}}
\caption{\label{compare} 
The total integrated intensity of Raman spectra divided by $\omega^4$ 
(see Eq.~(\ref{sumovers})) computed with
Placzek's approximation and with the approximation developed in this
paper, as a function of the incident laser beam energy. We present the intensity in a logarithmic scale. 
}
\end{figure}

Finally, in Fig.~\ref{compare}, we compare the total Raman intensities computed
using our method and the Placzek's approximation. The Raman
spectra computed with the Placzek's approximation show stronger
singularities in correspondence to the   $sp^2-sp^2$ electronic
transitions. Instead, the intensity of the spectra computed with our 
approach is similar for laser frequency in resonance with $sp^2-sp^2$
transitions and with $sp^3-sp^3$ transitions. We did not find any
experimental data for the total 
intensity as a function of the laser frequency, but according to the
discussion of Section \ref{theory} the use of our approximation is
more justified in the resonant case.

\section{conclusion}

In this paper, we have  presented a practical method to compute the
vibrational resonant Raman spectra in solids. 
Our approach is justified if the electronic excitations are delocalized.
We use our method to compute the vibrational resonant Raman spectra of
tetrahedral amorphous carbon.
The computed spectra are in good agreement with the experimental ones
measured with visible and UV lasers. 
In particular, the theoretical visible Raman spectra present a broad
feature between 1200 and 1700 cm$^{-1}$ associated with $sp^2$
stretching as found in experiments.\cite{} In the theoretical UV
Raman spectra a second broad peak around 1000 cm$^{-1}$ associated to
$sp^3$ stretching  shows up. The location and the shape of this peak
agree very well with the experimental UV spectra. \cite{}  
We have analyzed in detail the evolution of the Raman spectra as a
function of the 
laser frequency.  We have shown that, at any frequency, the spectra
are dominated by  
the stretching vibrations.
We have identified a very rapid inversion in the relative Raman intensities of
the $sp^2$ and $sp^3$ sites with the frequency of the incident laser beam.
In particular, the spectra are dominated by $sp^2$ atoms below 4 eV and
by $sp^3$ atoms above 6 eV.
According to our results, it would be interesting to collect a UV spectrum of 
tetrahedral amorphous carbon with a laser frequency larger than those used
in the actual experiments, to further emphasize the contribution of the 
tetrahedral $sp^3$ carbon sites.

We acknowledge D. A. Drabold for providing us with the model of
tetrahedral amorphous carbon.
We thank Dr. M. Marangolo for a critical reading of the manuscript.
The calculations have been performed at the Idris computer center
of the CNRS.

%\bibliography{ac}

\begin{thebibliography}{10}

\bibitem{Robertson}
A.~C. Ferrari and J. Robertson, Phys. Rev. B {\bf 61}, 14095 (2000), and
references therein.

\bibitem{visible}
Q. Wang, D.~D. Allred, and J. Gonz\'alez-Hern\'andez, Phys. Rev. B {\bf 47},
  6119  (1993).

\bibitem{uv}
V.~I. Merkulov {\it et~al.}, Phys. Rev. Letter {\bf 78},  4869  (1997).

\bibitem{uv2}
K.~W.~R. Gilkes {\it et~al.}, Appl. Phys. Lett. {\bf 70},  1980  (1997).

\bibitem{bruesh}
P. Br\"uesch, {\em Phonons, theory and experiments II : experiments and
  interpretation of experimental results} (Springer-Verlag, Berlin Heidelberg
  New York, 1986).                                                       

\bibitem{footnotehbar}
Here and in the following we will use atomic units, with which $\hbar=1$.

\bibitem{Placzek}
G. Placzek,  in {\em Handbuch der Radiologie}, edited by E. Marx (Akademische
  Verlagsgesellschaft, Leipzig, 1934), Vol.~6, p.\ 209.                 

\bibitem{porezag}
D. Porezag and M.~R. Pederson, Phys. Rev. B {\bf 54},  7830  (1996).

\bibitem{giannozzi}
P. Giannozzi and S. Baroni, J. Chem. Phys. {\bf 100},  8537  (1994).

\bibitem{cardona}
P.~V. Santos {\it et~al.}, Phys. Rev. B {\bf 52},  12158  (1995).

\bibitem{quantique}
R. Loudon, {\em The Quantum Theory of Light} (Clarendon Press ; Oxford
  University Press, New York, 1983).                                    



\bibitem{drabold}
D.~A. Drabold, P.~A. Fedders, and P. Stumm, Phys. Rev. B {\bf 49},  16415
  (1994).

\bibitem{mauriNMRac}
F. Mauri, B.~G. Pfrommer, and S.~G. Louie, Phys. Rev. Lett. {\bf 79},  2340
  (1997).

\bibitem{bernasconi}
N. Marks {\it et~al.}, Phys. Rev. B {\bf 54},  9703  (1996).

\bibitem{cpmd}
J. Hutter {\it et~al.}, CPMD Version 3.3.5, MPI f\"ur
Festk\"orperforschung and IBM Research Laboratory, 1990-1998.

\bibitem{tm2}
N. Troullier and J.~L. Martins, Phys. Rev. B {\bf 43},  1993  (1991).

\bibitem{kmho}
C.~H. Xu, C.~Z. Wang, C.~T. Chan, and K.~M. Ho, J. Phys. Condens. Matter {\bf
  4},  6047  (1992).

\bibitem{rpa}
L. X. Benedict, E. L. Shirley, and R. B. Bohn, Phys. Rev. B {\bf 57}, 
R9385 (1998).
%\bibitem{Calleja}
%J.~M. Calleja, J. Kuhl, and M. Cardona, Phys. Rev. B {\bf 17}, 876
%(1978).

\bibitem{footnotestreben}
To perform the projection we have defined for each bond a `stretching' vector
  in the space of the $3 N_{\rm atom}$ displacements. The components of each
  vector involve the displacement of two atoms in the direction of the bond
  and with opposite orientations. We use these vectors as a (non-orthonormal)
  basis of the stretching subspace. We define the bending subspace as the
  complement of the stretching subspace.

\bibitem{francesco}
F. Mauri and A.~D. Corso, Appl. Phys. Lett. {\bf 75},  644  (1999).

\bibitem{susceptibilite}
Z.~Y. Chen and J.~P. Zhao, J. Appl. Phys. {\bf 87},  4268  (2000).

\bibitem{overlap}
In the expression of intensity, Eq.~(\ref{sumovers}), ${\mathcal{A}}_s
={\mathcal{A}}_s^{\rm stretching}+{\mathcal{A}}_s^{\rm bending}$ where the 
contributions  
${\mathcal{A}}_s^{\rm stretching}$ and ${\mathcal{A}}_s^{\rm bending}$
are obtained  
substituting in Eq.~(\ref{kappa}) the vibrational eigenvector  ${\bf q}_s$ 
with its projections in the stretching and bending subspaces, respectively.
The intensity is then proportional to $|{\mathcal{A}}_s|^2 =
|{\mathcal{A}}_s^{\rm stretching}|^2 +|{\mathcal{A}}_s^{\rm bending}|^2 + 2
{\mathcal{A}}_s^{\rm stretching}{\mathcal{A}}_s^{\rm bending} $. 
The first two terms
correspond to the contributions of stretching and bending modes,
respectively, whereas the last term corresponds to the overlap term.
A similar repartition is used to define the $sp^3$ and $sp^2$ contributions.

\end{thebibliography}

\end{document}